\documentclass{ws-procs11x85}
\usepackage{balance}

\def\Journal#1#2#3#4{{#1} {\bf #2}, #3 (#4)}

\makeindex

\newcommand{\met}{\hbox{E\kern-0.5em\lower-0.1ex\hbox{/}}_T}

\begin{document}

\title{Statistical Computations with AstroGrid and the Grid}

\author{Robert Nichol}
\address{Institute of Cosmology and Gravitation (ICG), 
Univ. of Portsmouth, Portsmouth, PO1 2EG, UK}

\author{Garry Smith}
\address{ICG Portsmouth \& Institute of Astronomy, School of Physics, University of Edinburgh, UK.} 

\author{Christopher Miller}
\address{Cerro-Tololo Inter-American Observatory, NOAO, Casilla 603, LaSerena, Chile}

\author{Chris Genovese, Larry Wasserman}
\address{Dept. of Statistics, Carnegie Mellon University, Pittsburgh, PA-15213, USA}

\author{Brent Bryan, Alexander Gray, Jeff Schneider, Andrew Moore}
\address{School of Computer Science, Carnegie Mellon University, Pittsburgh, PA-15213, USA}


\twocolumn[\maketitle\begin{abstract} 
  
  We outline our first steps towards marrying two new and emerging
  technologies; the Virtual Observatory (e.g, AstroGrid) and the
  computational grid. We discuss the construction of {\it
    VOTechBroker}, which is a modular software tool designed to
  abstract the tasks of submission and management of a large number of
  computational jobs to a distributed computer system.  The broker
  will also interact with the AstroGrid workflow and MySpace
  environments. We present our planned usage of the {\it VOTechBroker}
  in computing a huge number of n--point correlation functions from
  the SDSS, as well as fitting over a million CMBfast models
  to the WMAP data.

\end{abstract}]

\bodymatter

\baselineskip=13.07pt
\section{Introduction}
Over a petabyte of raw astronomical data is expected to be collected
in the next decade (see Szalay \& Gray 2001).  This explosion of data
also extends to the volume of parameters measured from these data
including their errors, quality flags, weights and mask information.
Furthermore, these massive datasets facilitate more complex analyses,
e.g.  nonparametric statistics, which are computationally
intensive.  A key question therefore is: Can existing statistical
software scale-up to cope with such large datasets and massive
calculations? We address this question here.

We focus here on two exciting new technologies, namely the Virtual
Observatory (VO) and computational grids. However, we point the reader
to Jim Linnemann paper in these proceedings for an excellent summary
of existing statistical software packages in physics and astrophysics.
We also direct the reader to the recent ADASS conference proceedings
and the ``Mining the Sky'' proceedings
(www.mpa-garching.mpg.de/\~cosmo/).

\section{N--point Correlation Functions}

As a case study of the types of massive calculations planned for the
next generation of astronomical surveys and analyses, we discuss here
the galaxy n-point correlation functions. These have a long history in
cosmology and are used to statistically quantify the degree of spatial
clustering of a set of data points (e.g. galaxies). There are a
hierarchy of correlation functions, starting with the 2-point
correlation function, which measures the joint probability of a data
pair, as a function of their separation $r$, compared to a Poisson
distribution, i.e., $dP_{12} = N^2 dV_1\,dV_2 (1+\xi(r))$, where
$dP_{12}$ is the joint probability of an object being located in both
search volumes $dV_1$ \& $dV_2$, and $N$ is the space density of
objects. $\xi(r)$ is the 2-point correlation function
and is zero for a Poisson distribution. If $\xi(r)$ is positive, then
the objects are more clustered on scales of $r$ than expected, and
vice versa for negative values.

The next in the series is the 3-point correlation function, which is
defined as $dP_{123} = N^3 dV_1\,dV_2\,dV_3
(1+\xi_{12}(r_{12})+\xi_{23}(r_{23})+\xi_{13}(r_{13}) +
\xi_{123}(r_{12},r_{23},r_{13}) )$, where $\xi_{12},\xi_{12},\xi_{12}$
are the 2-point functions for the three sides ($r_{12},r_{23},r_{13}$)
of the triangle and $\xi_{123}$ is the 3--point function. Likewise,
one can define a 4-point, 5-point etc., correlation function. The
reader is referred to Peebles (1980) for a full discussion of these
n-point correlation functions including their importance to cosmology
(see also the recent lecture notes of Szapudi 2005). We also refer the
reader to Landy \& Szalay (1993) and Szapudi \& Szalay (1998) for a
discussion of the practical details of computing the N--point
functions.

Naively, the computation of the n--point correlation functions scale
as $O(R^n)$, where $R$ is the number of data--points in the sample. As
one can see, even with existing galaxy surveys from the Sloan Digital
Sky Survey (SDSS), where $R\sim 10^6$--$10^7$, such correlation
functions quickly become untractable to compute. In recent years,
there has been a number of more efficient algorithms developed to beat
this naive scaling. For example, the International Computational
Astrostatistics (inCA; www.incagroup.org) group has developed a new algorithm based on
the use of the multi--resolutional KD-tree data structure (mrKDtrees).
This software, known as {\it npt}, is publicly available
(www.autonlab.org/autonweb/software/10378.html), and has been
discussed previously in Gray et al. (2003), Nichol et al.  (2001) and
Moore et al. (2000). Briefly, mrKDtrees represent a condensed data
structure in memory, which is used to efficiently answer as much of
any data query as possible, i.e., pruning the tree in memory. The key
advance of our {\it npt} algorithm is the use of ``n'' trees in memory
together to compute an n--point function.  See also Alex Gray's
contribution in this volume.

\begin{figure}[t]
\centerline{\psfig{figure=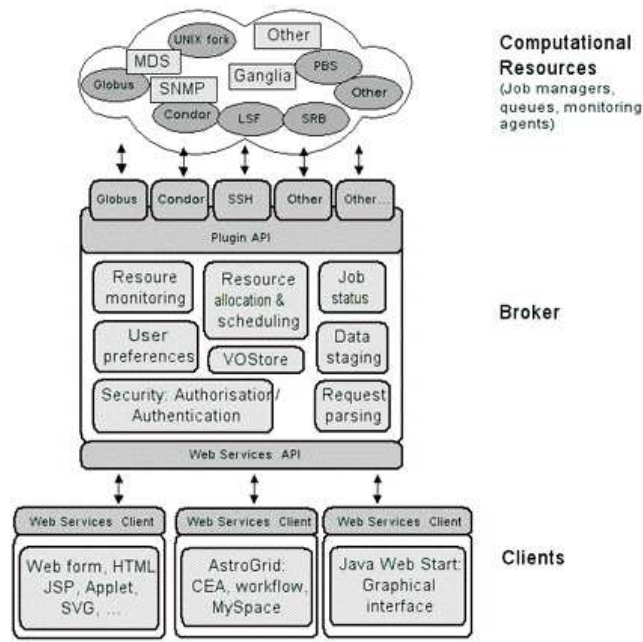,width=8.5cm}}
\caption{The archtecture of the VOTechBroker and how it interacts with the Grid, VO and our statistical algorithms. The {\it npt} algorithm is a ``Client'' (at the bottom) and interacts with the ``Broker'' via a web--form (HTML) to define the basic parameters needed to run the algorithm and define the resources needed. Eventually we plan to interact with the ``Broker'' via the AstroGrid workflow environment, allowing the submission of jobs as well as the storage of the input data and results in MySpace. There can be multiple ``Clients'' to the ``Broker''.}  \label{fig1}
\end{figure}

\section{Computing Correlation Functions}

Even with an efficient algorithm, the computation of higher--order
correlation functions is intensive.  In detail, the n--point
correlation functions require a large number of sequential calls to
the {\it npt} code. These include computing the cross--correlation
between the real data (called $D$) and a random dataset (called $R$),
which is used to mimic the edge effects in the real data. As outlined
in Szapudi \& Szalay (1998), each estimation of a 3--point correlation
functions, for a given bin of triangular shape (i.e.,
$r_{12}\pm\Delta_{r_{12}}$, $r_{23}\pm\Delta_{r_{23}}$,
$r_{13}\pm\Delta_{r_{13}}$, requires seven separate source counts over
the whole dataset, namely $DDD, DDR, DRR, RRR, DD, RR, DR$. Therefore,
if one wished to probe $\sim 10^2$ triangle configuration, then
$\sim10^3$ sequential {\it npt} jobs are required.  This can rise
rapidly if one wishes to estimate errors on the n--point functions
using either jack-knife resampling (i.e., removing subregions of the
data and then re-computing the correlation functions), or a large
ensemble of mock catalogs (derived from simulations). Such
computations are well-suited to large clusters or grid of computers.

In recent years, we have used computational resources like TeraGrid
(www.teragrid.org) and COSMOS (www.damtp.cam.ac.uk/cosmos/) to perform
the computation of the n--point correlation functions for the SDSS
main galaxy sample and the SDSS LRG sample. Our experience shows that
the management and scheduling of such a large number of jobs on these
massive machines is laborious and tedious. To ease this problem, we
are working on {\it VOTechBroker}, which is a tool that joins two new
and emerging technologies; the VO and computational grids.

\section{VOTechBroker}

AstroGrid (www.astrogrid.org) is a PPARC-funded project to create a
working Virtual Observatory for UK and international astronomers.
AstroGrid works closely with other VO initiatives around the world
(via the International Virtual Observatory Alliance; IVOA) and is part
of the Euro--VO initiative in Europe. In particular, the work outlined
here has been performed as part of the EU--funded VOTech project,
which aims to complete the technical preparation work for the
construction of a European Virtual Observatory. Specifically, VOTech
is undertaking R\&D into data--mining and
visualization tools, which can be integrated into the emerging VO and
computational grid infrastructure. Therefore, VOTech will build upon
existing or emerging standards and infrastructure (e.g. IVOA standards
and AstroGrid middleware), as well as looking at standards from W3C
and GGF.

As part of the VOTech research, we are engaged in developing the {\it
  VOTechBroker}. The key design goals of the broker are to: {\it i)}
Remove the execution and management of a large number of jobs (like
{\it npt}) from the user in a transparent and reusable way; {\it ii)}
Accommodate different grid infrastructures (e.g. condor, globus etc.);
{\it iii)} Locate suitable resources on the grid and optimize the
submission of jobs; {\it iv)}
Monitor the status and success of jobs; {\it v)} Combine with AstroGrid
MySpace and workflow environments to allow easy management of job
submission and final results (as well as utilizing other algorithms
within the VO). In Figure \ref{fig1}, we show the schematic design of
the broker archtecture which illustrates the modular and ``plug-in''
design philosophy we have adopted. This is required as one of the key
requirements of {\it VOTechBroker} is that it should be straightforward to
add new algorithms, resources and middleware (e.g. a different job
submission tool or protocol).

We have implemented the core functionality of {\it VOTechBroker} and
are presently testing it by submitting $\sim10^4$ {\it npt} jobs on
both the UK National Grid Servise (www.ngs.ac.uk), COSMOS
supercomputer and a local condor pool of machines. The key ingredients
of the present {\it VOTechBroker} include GridSAM (an open-source job
submission and monitoring web servise from the London e-Science
Centre), the UK e-Science X.509 certificates, MyProxy (a repository
for X.509 Public Key Infrastructure security credentials), and the Job
Submission Description Language (JSDL; a standard description of job
execution requirements to a range of resource managers from the Global
Grid Forum). At present, the {\it VOTechBroker} provides a web-form
interface to just the {\it npt} algorithm discussed above but is
modular in design so other algorithms can be easily added via other
web--forms. Results from the {\it VOTechBroker} will soon be placed in a user's
AstroGrid MySpace. In the near future, we will interface the broker
with other computational resources, e.g., TeraGrid (see below), and
the AstroGrid workflow.

\begin{figure}[t]
\centerline{\psfig{figure=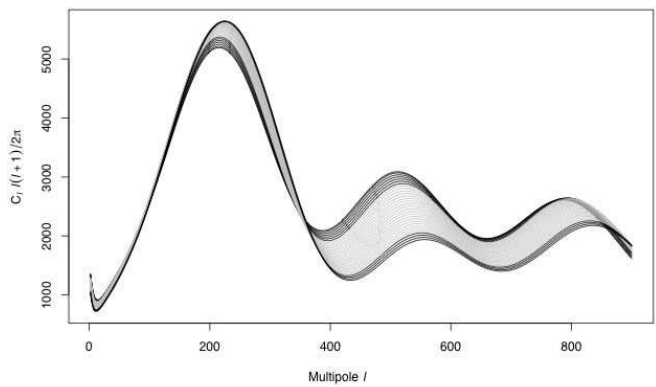,width=9cm}}
\caption{Using CMBfast, we have varied $\Omega_b$ (baryon fraction) and determined which models lie within the 95\% confidence ball around $f(X_i)$. For this illustration, we have kept all other parameters in these CMBfast models fixed at their fiducial values. The gray models are within the confidence ball, while the others are outside the ball indicating they are ``bad fits'' to the data (at the 95\% confidence). We get an allowed range of $0.0169<\Omega_b<0.0287$.\label{fig2}}
\end{figure}

\section{Nonparametric Statistics}

In addition to the need for new statistical software that scales-up to
petabyte datasets, we also require new algorithms and computational
resources that exploit the emerging power of nonparametric statistics.
As discussed in Wasserman et al. (2001), such nonparametric methods
are statistical techniques that make as few assumptions as possible
about the process that generated the data.  Such methods are more
flexible than more traditional parametric methods that impose rigid
and often unrealistic assumptions. With large sample sizes,
nonparametric methods make it possible to find subtle effects which
might otherwise be obscured by the assumptions built into parametric
methods.

In Genovese et al. (2004), we discuss the application of nonparametric
techniques to the analysis of the power spectrum of anisotropies in
the Cosmic Microwave Background (CMB).  For example, one can ask the
simple question: How many peaks are detected in the WMAP CMB power
spectrum? This question is hard to answer using parametric models for
the CMB (e.g. CMBfast models) as these models possess
multiple peaks and troughs, which could potentially be fit to noise
rather than real peaks in the data.  To solve this, we have performed
a nonparametric analysis of the WMAP power spectrum (Miller et al.
2003), which involves explaining the observed data ($Y_i$) as
$Y_i=f(X_i) + c_i$ where $f(X_i)$ is a orthogonal function (expanded
as a cosine basis $\beta_i{\rm cos}(i\pi X_i)$) and $c_i$ is the
covariance matrix. The challenge is to ``shrink'' $f(X_i)$ to keep the
number of coefficients ($\beta_i$) to a minimum. We achieve this using
the method of Beran (2000), where the number of coefficients kept is
equal to the number of data points. This is optimal for all smooth
functions and provides valid confidence intervals. We also use
monotonic shrinkage of $\beta_i$, specifically the nested subset
selection (NSS). The main advantage of this methodology is that it
provides a ``confidence ball'' (in N dimensions) around $f(X_i)$,
allowing non-parametric interferences like: Is the second peak in the
WMAP power spectrum detected? In addition, we can test parametric
models against the ``confidence ball'' thus quickly assessing the
validity of such models in N dimensions. This is illustrated in Figure
\ref{fig2}.

\section{Massive Model Testing}

We are embarked on a major effort to jointly search the 7--dimensional
cosmological parameter--space of $\Omega_m,\Omega_{DE},\Omega_b,\tau$,
neutrino fraction, spectral index and H$_0$ using parametric models
created by CMBfast and thus determine which of these models fit within
the confidence ball around our $f(X_i)$ at the 95\% confidence limit.
Traditionally, this is done by marginalising over the other parameters
to gain confidence intervals on each parameter separately. This is a
problem in high-dimensions where the likelihood function can be
degenerate, ill-defined and under-identified. Unfortunately, the
nonparametric approach is computational intense as millions of models
need to searched, each of which takes $\simeq3$ minute to run.

\begin{figure*}
\centerline{\psfig{figure=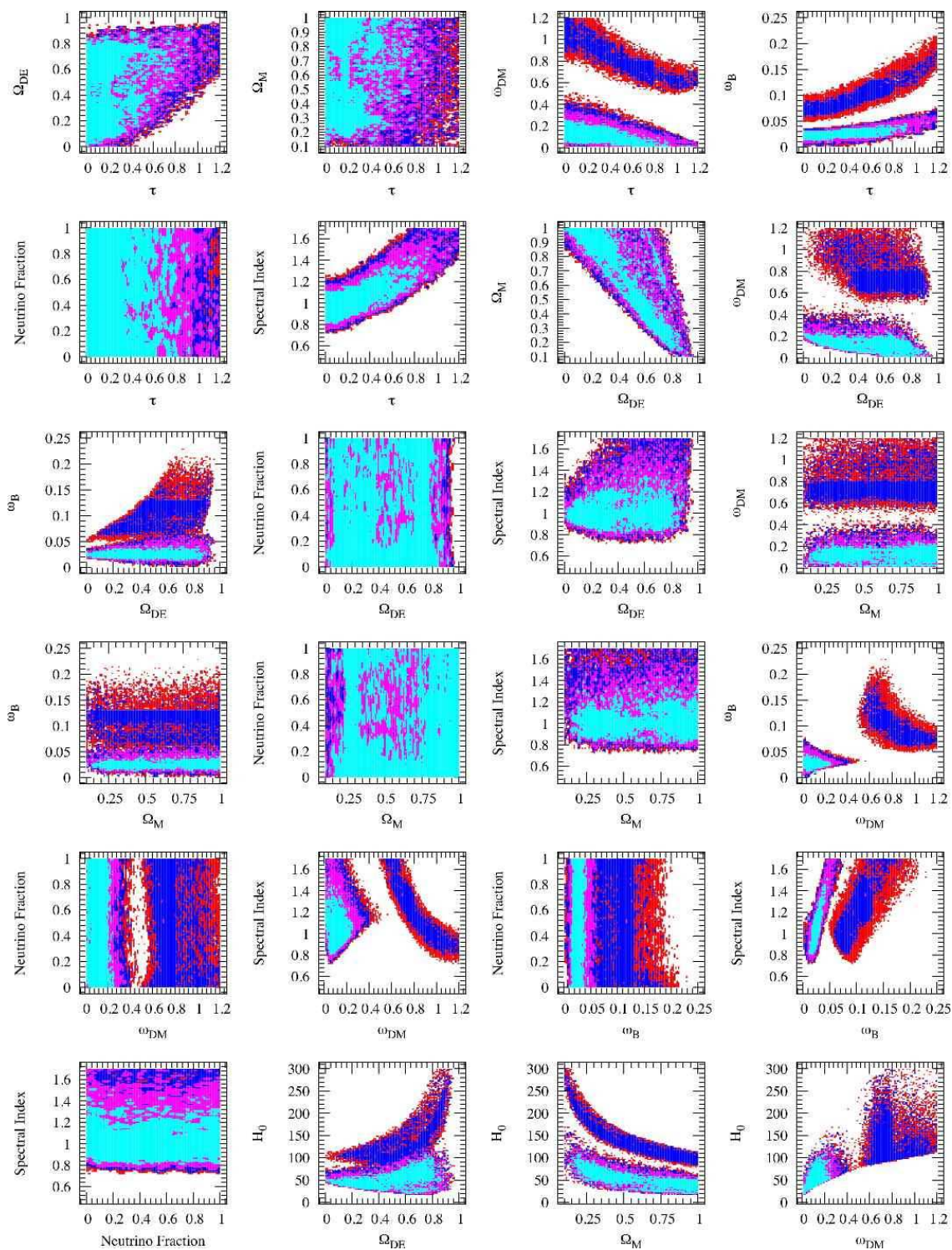,width=15cm}}
\caption{The results of our 7--dimensional parameter search using 1.2 million models from CMBfast. The light blue (or lightest shading for greyscales) color are models excluded at the 34\% level. The purple (or mid-grade shading) are models excluded by the 68\% confidence ball and the red is the 95\% confidence ball \label{fig3}}
\end{figure*}

To mitgate this problem, we have developed an intelligent method for
searching for the surface of the confidence ball in high-dimensions
based on Kriging. Briefly, kriging is a method of interpolation which
predicts unknown values from data observed at known locations (also
known as Gaussian process regression, which is a form of Bayesian
inference in Statistics). There are many different metrics for
evaluating the kriging success; we use here the ``Straddle'' method
which picks new test points based both on the overall distance from
previous searched points, as well as being predicted to be close to
the boundary of the confidence ball. We have also developed a
heuristic algorithm for searching for ``missed peaks'' in the
likelihood space by searching models along the path joining previously
detected peaks. We find no ``missed peaks'', which illustrates our
kriging algorithm is effective in finding the surface of the
confidence ball in this high dimensional space.

We have distributed the CMBfast model computations over a local condor
pool of computers. In Figure \ref{fig3}, we show preliminary results
from this high-dimension search for the surface of the confidence ball
and present {\bf joint} 2D confidence limits on pairs of the
aforementioned cosmological parameters. These calculations represent
6.8 years of CPU time to calculate over one million CMBfast models. In
the near future, we will move this analysis to TeraGrid, using {\it
  VOTechBroker}, and plan 10 million models to fully map the surface
of the confidence ball. We will also make available a Java--based web
servise for accessing these models, and the WMAP confidence ball, thus
allowing other users to rapidly combine their data with our WMAP
constraints e.g., doing a joint constraint from LSS and CMB data. We
are also working on possible convergence tests, and visualization tools
within VOTech, to access this high-dimensional data.

\section{Summary}

The two examples given here -- massive model testing of the WMAP data
using nonparametric statistics and higher--order correlation functions
of SDSS galaxies -- represent a growing trend in astrophysics and
cosmology for massive statistical computations. Our plan is to develop
the {\it VOTechBroker} to provide a power framework within which such
massive astronomical analyse can be performed. As discussed, the main
goals of the {\it VOTechBroker} are to abstract from the user (either
a person or another program) the complexities of job submission and
management on computational grids, as well as being a modular
``plug--in'' design so other algorithms and software can be easily
added. Finally, we plan to integrate {\it VOTechBroker} into the
AstroGrid workflow and MySpace environments, so it becomes a natural
repository for a host of advanced statistical algorithms than scale-up
in preparation for petabyte-scale datasets and analyses.

\section*{Acknowledgments}

We thank all my collaborators and colleagues in inCA, VOTech,
AstroGrid, SDSS and VO projects. The work presented here was partly
funded by NSF ITR Grant 0121671 and through the EU VOTech and Marie
Curie programs. RCN thanks the organisers of the Phystat2005 meeting
for their invitation. GS thanks the VOTech and University of Edinburgh
for his funding (see eurovotech.org for details).

\balance

\end{document}